# How Do Data Collection Strategy and Data Quality Influence the Outcomes of Digital Technology Adoption?

Xuejiao Li[1*], Yang Cheng[2]
1: Center for Supply Chain Digitalisation, Department of Technology and Innovation, University of Southern Denmark, Denmark
2: Department of Materials and Production, Aalborg University, Denmark
* Corresponding author: xuejiao@iti.sdu.dk

## Abstract

In the era of Industry 4.0 (I4.0), data has become the essential foundation for digital transformation, yet many organizations still struggle to link data practices with digital performance outcomes. This study investigates how data collection strategy and data quality jointly influence the success of digital technology adoption (DTA) in manufacturing firms. Drawing on survey data from 86 firms, the research employs Partial Least Squares Structural Equation Modeling (PLS-SEM) to examine the relationships among data collection strategy, data quality, implementation performance, and operational performance. The results show that both data collection strategy and data quality significantly influence implementation and operational performance. However, the effect of data collection strategy on implementation performance is indirect, fully mediated by data quality. The study also finds that data collection strategy influences data quality. These findings demonstrate that data quality acts as a critical bridge between upstream data practices and downstream digital outcomes. The study contributes to the digital transformation and data management literature by empirically validating the central role of data quality and offering practical insights for managers to design and govern data processes strategically. It also sets a foundation for future research on data governance frameworks that integrate data quality assurance, standardization, and lifecycle management to sustain data-driven and digital transformation.

**Keywords:**
Data collection strategy; Data quality; Digital technology adoption (DTA); Implementation performance; Operational performance; PLS-SEM

## 1. Introduction

In the era of Industry 4.0 (I4.0), digital technologies have emerged as pivotal enablers of digital transformation across industries, particularly in operations and supply chain management (OSCM). From predictive analytics and machine learning (ML) to advanced planning systems and Internet of Things (IoT) applications, these technologies promise substantial gains in efficiency, agility, and competitiveness (Lee et al., 2015; Zonta et al., 2020). However, the success of such digital initiatives depends not only on the technologies themselves but critically on the availability and quality of data that underpins them. In many organizations, digital technology investments have failed due to inadequate attention to data, the foundational element of digital technology adoption (DTA) (Bosu & MacDonell, 2013; Ramesh & Delen, 2021; Esther, 2024). Drawing from practical experience and industry observations, it has become increasingly evident that organizations often overlook the importance data. As data has become the "raw material" of digital transformation (Möller, 2020), understanding how data is collected, managed, and utilized is crucial for ensuring meaningful and sustainable digital outcomes.

While the literature on digital transformation, I 4.0, and data-driven decision-making is vast and growing, existing research tends to focus on the organizational or strategic aspects of successful DTA (e.g., Kumar et al., 2021; Marcon et al., 2022; Kiraz et al., 2020; Tortorella et al., 2022). Far less attention has been given to data-related enablers, particularly the upstream processes that determine data usability in digital contexts. Data can be viewed across multiple dimensions, yet two aspects emerge as particularly foundational: data collection strategy and data quality. Data collection strategy defines how and what data are gathered, from which sources, and under what level of standardization and control (Bilsborrow, 2016). It determines the structure, completeness, and relevance of the data that digital technologies rely upon. Data quality, in turn, reflects the degree to which data are accurate, consistent, timely, and reliable (Wang & Strong, 1996; Evans, 2006). Despite their importance, most studies treat data merely as an assumed input rather than as a design variable, rarely examining how data collection strategies or data quality influences DTA outcomes. This reveals a clear research gap: the lack of systematic, empirical investigation into how data collection strategy and data quality jointly affect the effectiveness and performance outcomes of DTA.

The outcomes of DTA can be viewed from both implementation and post-adoption perspectives. During implementation, performance depends on whether digital initiatives are completed on time, within budget, and according to expectations. After implementation, other indicators, such as cost efficiency, flexibility, delivery reliability, sales growth, and profitability, capture the broader impact of digital technologies on companies' performance. Studying these two layers together provides a more comprehensive understanding of how digital transformation unfolds from project-level execution to organization-level outcomes. However, limited research has empirically linked upstream data practices to both implementation success and operational performance in digital adoption contexts, especially using quantitative models.

Motivated by these gaps, this paper seeks to understand how data collection strategy and data quality jointly influence the outcomes of DTA. Specifically, it aims to assess their effects on the implementation performance of DTA projects, and the operational performance of companies after they have implemented digital technologies. This study also examines the interrelationship between data collection strategy and data quality. To achieve these goals, the study explores the following three research questions:

**RQ1:** How does data collection strategy influence implementation performance and operational and performance in the context of DTA?

**RQ2:** How does data quality influence implementation performance and operational and performance in the context of DTA?

**RQ3:** How does data collection strategy influence data quality?

To investigate these questions, we conducted a structured survey of 86 manufacturing firms. The survey measured firms' data collection practices, perceived data quality, and performance outcomes related to digital technology implementation and operation. All constructs were measured with reflective indicators on five-point Likert scales, adapted from validated studies. The data were analyzed using Partial Least Squares Structural Equation Modeling (PLS-SEM), an approach particularly suited for exploratory studies with complex, multi-dimensional models and modest sample sizes.

In doing so, this study makes several contributions. Theoretically, it extends digital transformation and data governance literature by empirically validating the links between data collection strategy, data quality, and performance outcomes. It advances understanding of how upstream data practices shape downstream digital success, emphasizing that data should not be treated as a passive input but as a strategic asset that determines digital effectiveness. Practically, the study offers evidence-based insights into how data collection strategy and data quality impact digital outcomes. It highlights that merely investing in advanced technologies is insufficient without a robust foundation of high-quality, systematically collected data. Managers can use these insights to strengthen digital project planning and governance, ensuring that data collection and management processes are aligned with performance objectives from the outset.

The remainder of this paper is organized as follows. Section 2 reviews relevant literature and develops the conceptual framework and hypotheses. Section 3 describes the research methodology, including survey design and PLS-SEM modeling. Section 4 presents empirical results and interprets findings. Section 5 discusses theoretical and managerial implications, and Section 6 concludes the study, and points out the limitations and future research directions.

## 2. Literature Review and Hypothesis Development

### 2.1 DTA and its Outcomes

DTA refers to the organizational integration of technologies such as big data analytics, artificial intelligence (AI), IoT, blockchain, and advanced planning systems into core operations and strategic practices (Yang et al., 2021; Skare & Soriano, 2021; Blichfeldt & Faullant, 2021). Particularly in manufacturing and OSCM, these technologies are crucial for improving decision-making, enhancing responsiveness, reducing waste, and sustaining competitive advantage. Successful DTA has become a key driver of firm long-term performance.

The outcomes of DTA can be observed across two interrelated dimensions: implementation performance, i.e., the extent to which digital initiatives are successfully deployed and integrated into existing processes; and operational performance, which captures the efficiency, resilience, and sustainability outcomes enabled by these technologies.

From an implementation perspective, research has shown that the success of DTA depends on how effectively digital technologies are integrated into organizational workflows, supported by appropriate structures, skills, and data capabilities. Martínez-Caro et al. (2020) emphasized that digital investments yield meaningful results only when backed by a digital-oriented culture that facilitates adoption and learning. Tsou et al. (2021) similarly found that digital strategy alignment and organizational innovation serve as critical enablers linking technology implementation to improved operational outcomes. Studies of small and medium-sized enterprises (SMEs) also confirm this pattern. Iskandar et al. (2023) reported that firms adopting digital tools improved internal productivity and service delivery, illustrating that effective implementation is central to realizing digitalization benefits. Collectively, these findings underline that DTA success is contingent not merely on technology acquisition, but on the organization's capability to implement and embed these tools in a coherent and scalable way.

Beyond implementation, a substantial body of research highlights the influence of DTA on companies' operational performance after implementing digital technologies, particularly in terms of efficiency, resilience, and sustainability. Li et al. (2020) showed that Industry 4.0 technologies enhanced environmental and economic outcomes in supply chains by enabling data-driven platforms. In the healthcare sector, Laurenza et al. (2018) found that digital adoption improved efficiency by reducing response times and improving service delivery. Zhai et al. (2022) also observed that digital transformation contributes to cost reduction and efficiency gains, while simultaneously fostering innovation. Furthermore, Zhou et al. (2024) demonstrated that digital technologies enhance supply chain resilience by lowering dependency on concentrated sources, and Wan et al. (2023) revealed that digital development in high-tech industries increased both efficiency and innovation, albeit unevenly across regions and firms. Together, these studies show that operational outcomes, including efficiency, resilience, and sustainability, are equally crucial benefits of DTA.

Overall, the literature demonstrates that DTA drives tangible operational and implementation benefits across industries. However, despite growing attention to digital transformation outcomes, empirical understanding remains fragmented. Most studies examine either the implementation process or operational results in isolation, leaving limited insight into how effective implementation practices translate into sustained operational performance improvements. Addressing this gap is crucial for developing a more comprehensive understanding of how digital adoption delivers value in practice, particularly within data-intensive operational environments.

### 2.2 The Enablers of DTA

Successful DTA in the time of I 4.0 depends on a variety of organizational and technical factors. Prior studies have emphasized the importance of leadership, organizational culture, technological infrastructure, external support, and data-related capabilities in determining the success across sectors and geographies.

#### 2.2.1 Leadership and Strategic Commitment

Several studies have underscored the pivotal role of leadership in enabling digital adoption. Arumugam et al. (2022) emphasized that leadership effectiveness significantly influences DTA in the electrical and electronic manufacturing sector, serving not only as a direct enabler but also as a mediator that amplifies the effect of other organizational dynamics. Similarly, Arumugam et al. (2023) expanded this insight through a conceptual framework grounded in TAM and DOI theories, arguing that leadership style and psychological traits shape an organization's openness to innovation. Top management commitment is also found to be essential for aligning digital strategy with long-term business objectives. Kumar et al. (2022) listed it as a critical driver in their empirical model of Industry 4.0 adoption for sustainability, alongside environmental regulations and stakeholder engagement. Kumar et al. (2021) further identified "top management commitment" and "knowledge of I4.0 and circular supply chains" as central to sustainable integration of digital technologies.

#### 2.2.2 Organizational and Sociotechnical Factors

Marcon et al. (2022) introduced a sociotechnical perspective to I4.0 adoption, showing how the alignment of social, technical, and organizational systems enhances digital transformation success. They found that work organization, team collaboration, and environmental preparedness contribute meaningfully to high adoption levels in Danish manufacturing firms. Kiraz et al. (2020) analyzed nine factors using SEM and concluded that customer access, value chain optimization, IT infrastructure, and data integration are the most influential criteria shaping I4.0 readiness. Similarly,

Tortorella et al. (2022) demonstrated that manufacturing strategy mediates the effect of critical success factors (CSFs) on technology adoption, particularly under high I4.0 readiness conditions. Shahadat et al. (2023), applying the TOE and DOI frameworks in emerging economies, found that relative advantage, complexity, cost, innovativeness of top management, and government support significantly influence ICT adoption among SMEs, revealing how external and internal forces intertwine in shaping readiness.

### 2.2.3 Innovation, Culture, and Psychological Factors

The importance of organizational culture and psychological readiness has been increasingly recognized. Gerli et al. (2022) contributed a novel perspective by integrating emotional and cognitive dimensions into the TAM framework. They showed that beliefs, learning attitudes, and emotional responses significantly influence skill development and the decision to adopt smart technologies in agriculture. Narula et al. (2020) also placed "leadership, strategy, and culture" at the heart of digital transformation, identifying these elements as essential enablers for the adoption of technologies like digital twins and virtual testing in manufacturing. Arbaiza (2018) identified critical success variables for ICT-based digital transformation using a cause-effect approach. The study emphasized that without identifying and addressing the specific variables influencing a project, digital efforts are likely to fall short. These include change readiness, communication, and continuous stakeholder engagement.

### 2.2.4 External and Institutional Support

Multiple studies point to the importance of institutional and policy environments in enabling digital transformation. Cagliano et al. (2021) used ANOVA to demonstrate that economic factors, such as GDP per capita, R&D investment, and FDI, influence the uptake of digital supply chain technologies. Their findings suggest that contextual factors are just as influential as internal readiness. Sichoongwe (2023), studying South Africa's manufacturing sector, found that innovation, firm size, infrastructure, and export activity influence DTA. The study revealed that adoption behaviors vary across business functions and are heavily influenced by persistence and prior exposure. Ramesh and Delen (2021), based on three case studies, proposed five enablers of digital transformation success: innovation attributes, opinion leaders, diffusion approaches, timing, and duration. These findings offer insights into why even well-resourced digital initiatives often fail due to poor orchestration and misaligned change management strategies.

### 2.2.5 Data Governance, Infrastructure, and Legal Context

Bhatia and Kumar (2020) identified "data governance" as the most critical enabler of I4.0 adoption in the Indian automotive sector, followed closely by legal and regulatory aspects. Their study demonstrated how data governance influences all performance outcomes, i.e., operational, product, economic, and responsiveness, underscoring the foundational importance of managing data as a strategic asset. Likewise, Aldossari et al. (2023) conducted a systematic literature review on big data analytics (BDA) adoption in SMEs and found that data-related enablers such as IT infrastructure, security, and data governance, along with top management support and training, are core elements for successful implementation. Their findings further stress the need for robust governance frameworks that support digital initiatives. Table 1 summarizes the enablers of successful DTA.

Table 1. The enablers of successful DTA

| Enabler Category | Key Factors | Sources |
|---|---|---|
| **1. Leadership and Strategic Commitment** | - Leadership effectiveness<br>- Leadership style and psychological traits<br>- Top management commitment<br>- Strategic alignment with digital goals | Arumugam et al. (2022, 2023);<br>Kumar et al. (2021, 2022) |
| **2. Organizational and Sociotechnical Factors** | - Work organization and team collaboration<br>- Environmental preparedness<br>- Value chain optimization<br>- Customer access<br>- IT infrastructure<br>- Manufacturing strategy | Marcon et al. (2022);<br>Kiraz et al. (2020);<br>Tortorella et al. (2022);<br>Shahadat et al. (2023) |
| **3. Innovation, Culture, and Psychological Factors** | - Organizational culture<br>- Emotional and cognitive readiness<br>- Change readiness<br>- Learning attitudes<br>- Communication and stakeholder engagement | Gerli et al. (2022);<br>Narula et al. (2020);<br>Arbaiza (2018) |
| **4. External and Institutional Support** | - Government and policy support<br>- Economic development factors (GDP, R&D investment, FDI)<br>- Digital infrastructure<br>- Firm innovation and export activity<br>- Timing and diffusion approach | Cagliano et al. (2021); Sichoongwe (2023);<br>Ramesh & Delen (2021) |
| **5. Data Governance, Infrastructure, and Legal Context** | - Data governance<br>- IT infrastructure<br>- Security and data integration<br>- Legal and regulatory compliance<br>- Training and technical support | Bhatia & Kumar (2020);<br>Aldossari et al. (2023) |

### 2.3 Data Collection Strategy, Data Quality, and the Missing Link in DTA Research

While previous studies have extensively explored leadership, culture, infrastructure, and institutional factors as enablers of DTA, relatively limited attention has been given to data-related foundations that underpin these processes. Digital transformation inherently depends on data. However, despite growing recognition of data as a strategic resource, empirical research on how data affects the outcomes of DTA remains scarce.

Data can be understood as a multifaceted resource encompassing aspects such as data governance, architecture, storage, privacy, and analytics capabilities. Among these dimensions, two stand out as particularly fundamental to DTA: data collection strategy and data quality. Data collection determines what information enters the organization's digital systems, while data quality determines how reliable and usable that information is for decision-making and automation. Inadequate data collection practices may lead to incomplete, inconsistent, or biased data streams, which in turn compromise data quality and limit the effectiveness of digital tools. Thus, focusing on these two dimensions provides a critical lens to understand the mechanisms through which data-related factors shape digital adoption outcomes.

Data collection strategy refers to the systematic approach an organization uses to gather, store, and organize data from various sources, such as sensors, enterprise systems, customer interactions, or external platforms (Sivarajah et al., 2017). The choice of data collection methods influences the accuracy, consistency, and timeliness of data available for digital tools. Studies have shown that effective data acquisition and integration are crucial for analytics and automation performance (Li et al., 2018; Wang et al., 2018). Yet, most research discusses data collection only implicitly, often subsuming it under broader categories such as IT infrastructure, data readiness, or integration capability. For instance, Kiraz et al. (2020) and Marcon et al. (2022) acknowledged "data integration" and "IT infrastructure" as important readiness factors for digital transformation, they did not examine how the design and management of data collection processes affect the success of DTA. This highlights a clear conceptual and empirical gap: the impact of data collection strategy on DTA remains underexplored, limiting our understanding of how data acquisition practices shape implementation success and performance outcomes.

Data quality, in turn, determines whether collected data is accurate, complete, consistent, and reliable enough to support digital operations (Batini et al., 2016). Poor data quality can significantly undermine the potential of digital technologies, leading to biased models, inaccurate analytics, and misguided decisions (Hazen et al., 2014; Abbasi et al., 2016). Although some studies have emphasized the role of data quality in specific domains—such as big data analytics (Wamba et al., 2017) and supply chain management (Akyuz & Rehan, 2009), its relationship with DTA outcomes has rarely been explored systematically. Most frameworks account for technological and organizational readiness but treat data quality as a secondary technical issue rather than a key determinant of adoption success (Tortorella et al., 2022). Moreover, the interdependence between data collection strategy and data quality has received little attention, even though they are closely linked. In practice, an organization's approach to collecting, integrating, and managing data directly affects data quality, which in turn shapes the reliability and effectiveness of digital systems (Fan & Geerts, 2012; Otto, 2011).

In sum, while the literature has made significant progress in identifying organizational, cultural, and infrastructural enablers of digital adoption, the data foundation remains underexamined. There is a pressing need to understand how data collection strategy and data quality jointly influence the implementation and operational outcomes of DTA. Addressing this gap is essential because the success of digital transformation depends not only on adopting advanced technologies but also on managing the quality and structure of the data that fuel them.

## 2.4 Conceptual Framework and Hypotheses Development

Building on the reviewed literature, this study proposes a conceptual framework linking data collection strategy, data quality, and two key outcomes of DTA: implementation performance and operational performance. The framework is grounded in the premise that effective digital transformation depends not only on technological or managerial readiness but also on the quality of underlying data processes.

A well-designed data collection strategy defines how and what data are gathered, integrated, and standardized across digital systems (Sivarajah et al., 2017; Fan & Geerts, 2012). It establishes clear protocols regarding data sources, formats, frequency, and validation mechanisms. Such standardization ensures that only relevant, timely, and structured data enter organizational systems, creating a stable foundation for analytics, automation, and decision-making. When data collection is planned and systematic, it minimizes redundancies and inconsistencies, improving the traceability and reliability of data across departments. Conversely, ad hoc or fragmented data collection often results in missing, inconsistent, or duplicated records, which compromise data reliability and create bottlenecks for digital implementation (Batini et al., 2016; Otto, 2011).

Hence, data collection strategy is a determinant of data quality. By shaping the procedures and standards through which data are acquired, it directly influences whether data are accurate, complete, consistent, and timely—key dimensions of data quality (Wang & Strong, 1996). Studies have shown that standardized data acquisition enhances data integrity and usability (Li et al., 2018; Wang et al., 2018), while uncoordinated collection processes lead to poor-quality data and limited analytical value (Hazen et al., 2014). In this sense, data collection strategy functions as an upstream control mechanism, ensuring that high-quality data flow through the organization's digital infrastructure.

The quality of data, in turn, plays a mediating role between data collection and DTA performance. During digital implementation, high-quality data reduce system errors, enhance model accuracy, and ensure that digital tools such as predictive analytics or IoT platforms operate effectively. This facilitates smoother integration and project execution, improving implementation performance (Abbasi et al., 2016). Similarly, in day-to-day operations, high-quality data enable reliable monitoring, informed decision-making, and efficient resource allocation, leading to improved operational performance (Hazen et al., 2014).

Therefore, while data collection strategy sets the structural foundation, data quality determines whether that foundation translates into successful digital outcomes. Together, these two constructs form the data backbone of digital transformation.

The conceptual relationships among data collection strategy, data quality, and the performance outcomes of DTA are illustrated in Figure 1.

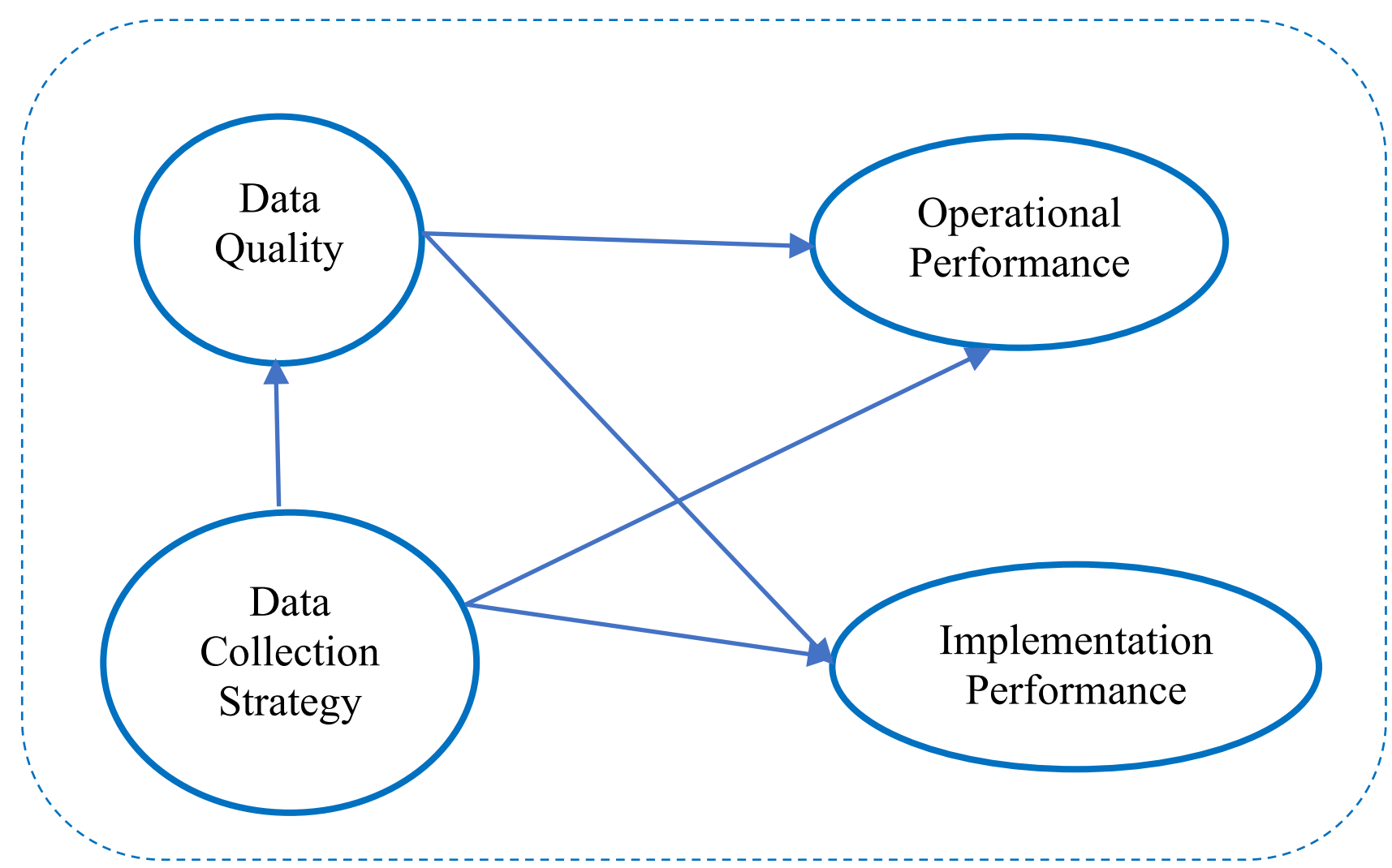

Figure 1. Conceptual Framework of the Study

Based on the above reasoning, the following hypotheses are proposed:

- H1: Data collection strategy positively influences implementation performance.
- H2: Data collection strategy positively influences operational performance.
- H3: Data collection strategy positively influences data quality.
- H4: Data quality positively influences implementation performance.
- H5: Data quality positively influences operational performance.
- H6: Data quality mediates the relationship between data collection strategy and implementation performance.
- H7: Data quality mediates the relationship between data collection strategy and operational performance.

## 3. Research Methodology

### 3.1 Research Design and Approach

Building upon the conceptual gaps identified in the literature review, this study empirically investigates how data collection strategy and data quality influence the outcomes of DTA. Specifically, a quantitative, survey-based approach was adopted. All constructs were measured using multi-item reflective indicators on a five-point Likert scale, where higher values indicated stronger implementation, coordination, or performance. More details about the items for each construct are presented in the questionnaire in the Appendix. Furthermore, this study applies Partial Least Squares Structural Equation Modeling (PLS-SEM) for hypothesis testing. PLS-SEM is particularly suitable for this study because it accommodates small to medium sample sizes and allows for complex models with both reflective and mediating constructs (Hair et al., 2017; Reinartz et al., 2009).

### 3.2 Measurement of Constructs

Following the conceptualization in Section 2, this study operationalizes four key constructs—data collection strategy, data quality, implementation performance, and operational performance—to empirically test the proposed relationships in the conceptual framework. All constructs were measured reflectively using multi-item indicators adapted from prior studies and assessed on five-point Likert scales.

*Data Collection Strategy*

Data collection strategy refers to the degree of formalization, standardization, and definition of data collection processes within an organization prior to digital implementation. It captures the degree to which firms formalize and standardize their data collection processes before implementing digital technologies. This includes defining specific data types and formats to be collected and ensuring that data collection procedures follow standardized workflows across business functions. A well-designed data collection strategy ensures that data entering digital systems are relevant, reliable, and fit for purpose, thereby forming the foundation for data quality and successful DTA (Bilsborrow, 2016; Fan & Geerts, 2012; Sivarajah et al., 2017). In this study, it was operationalized as a two-item construct:

- DP1: Standardization of business processes and clear definition of data types and formats.
- DP2: Standardization of the data collection process itself, ensuring consistent data acquisition across activities.

*Data Quality*

Data quality has been widely recognized as a multidimensional construct encompassing the degree to which data are accurate, timely, complete, consistent, interpretable, accessible, and relevant to their intended use (Ballou & Pazer, 1985; Wang & Strong, 1996; Evans, 2006). High-quality data reduce uncertainty in digital operations and enable effective analytics, automation, and decision-making. Wang and Strong (1996) divided data quality into four categories: intrinsic data quality, contextual data quality, representational data quality, and accessibility data quality. The dimensions and definitions of these four data qualities are shown in Table 2.

Table2. Dimensions and definitions of these four data qualities (Wang & Strong, 1996)

| Dimension | | Definition |
|---|---|---|
| **Intrinsic data quality** | Accuracy | The extent to which data is correct, reliable, and certified free of error. |
| | Believability | The extent to which data is accepted or regarded as true, real, and credible. |
| | Objectivity | The extent to which data is unbiased (unprejudiced) and impartial. |
| | Reputation | The extent to which data is trusted or highly regarded in terms of their source or content. |
| **Contextual data quality** | Value-added | The extent to which data is beneficial and provides advantages from their use. |
| | Relevancy | The extent to which data is applicable and helpful for the task at hand. |
| | Timeliness | The extent to which the age of the data is appropriate for the task at hand. |
| | Completeness | The extent to which data is of sufficient breadth, depth, and scope for the task at hand. |
| | Appropriate amount of data | The extent to which the quantity or volume of available data is appropriate. |
| **Representational data quality** | Interpretability | The extent to which data is in appropriate language and units and the data definitions are clear. |
| | Ease of understanding | The extent to which data is clear without ambiguity and easily comprehended. |
| | Representational consistency | The extent to which data is always presented in the same format and is compatible with previous data. |
| | Concise representation | The extent to which data is compactly represented without being overwhelming (i.e., brief in presentation, yet complete and to the point). |
| **Accessibility data quality** | Accessibility | The extent to which data is available or easily and quickly retrievable. |
| | Access security | The extent to which access to data can be restricted and hence kept secure. |

In this study, considering the context of manufacturing industry, we select accuracy (DQ1), timeliness (DQ2), completeness (DQ3), interpretability (DQ4), representational consistency (DQ5), accessibility (DQ6), and relevancy (DQ7) as measure items, consistent with existing literature (Ballou & Pazer, 1985; Wang & Strong, 1996; Evans, 2006). These seven indicators best capture the dimensions of data quality that are most relevant to the manufacturing and digital transformation context. Specifically, accuracy ensures that data correctly represents real-world conditions, which is critical for reliable decision-making in production and quality control; timeliness guarantees that data reflects the current operational state, enabling responsive adjustments; completeness ensures that data has sufficient breadth and depth to support comprehensive analysis; interpretability emphasizes the clarity of definitions and formats, facilitating understanding and system integration; representational consistency ensures uniform data presentation across systems and over time, supporting interoperability; accessibility emphasizes the ease and speed of data retrieval and sharing, which underpins data-driven decision-making; and relevancy ensures that data is applicable and useful for specific operational tasks. Together, these seven attributes capture the essential characteristics that determine how effectively manufacturing firms can leverage data for digital transformation, balancing theoretical comprehensiveness and practical applicability.

*Implementation Performance*

Implementation performance reflects the success of digital technology projects in terms of meeting planned objectives related to time, budget, functionality, and satisfaction. It captures whether the digital system was implemented efficiently and effectively, serving as an immediate indicator of adoption success (Chien et al., 2007; Ram et al., 2013). The construct is measured using four items:

- IP1: The digital technology implementation was completed on time.
- IP2: The implementation was completed within the planned budget.
- IP3: The implemented technology functioned as expected.
- IP4: The organization is satisfied with the implemented digital technology.

*Operational Performance*

Operational performance represents the downstream effects of digital technology adoption on firms' internal operations, including efficiency, quality, flexibility, and delivery reliability. It captures the extent to which digital adoption improves core operational outcomes that contribute to long-term competitiveness (Cheng et al., 2020; Li et al., 2020). Based on well-established dimensions of manufacturing performance, operational performance was assessed across four areas: cost, quality, flexibility, and delivery, and each represented by two items:

- Cost: Ordering cost (OPC1); Unit production cost (OPC2).
- Quality: Conformance quality (OPQ1); Product quality and reliability (OPQ2).
- Flexibility: Volume flexibility (OPF1); Mix flexibility (OPF2).
- Delivery: Delivery speed (OPD1); Delivery reliability (OPD2).

Together, these constructs provide a comprehensive measurement framework to examine how data collection strategy and data quality jointly shape both the implementation and operational performance outcomes of digital technology adoption.

### 3.3 Sampling and data collection

In this paper, we design a questionnaire to collect data from eighty-six manufacturing companies. The data were collected from November 2020 to February 2021. The positions of the respondents include CEO, CIO, IT manager, finance manager, procurement director, product manager, supply chain business analyst manager, planning manager, marketing manager, R&D manager, operations manager, chief engineer, etc. The respondents are all middle level or above managers in their companies, which ensures that they have a broad understanding of their companies' business. In addition, the diversity of their professions ensures that the views we collected are comprehensive.

The eighty-six companies are located in thirteen Chinese provinces and cities, as showing in Figure 2. Our survey covers both small and medium-sized companies and large companies. Of these companies, thirteen have 20 to 299 employees, accounting for 15% of the total, fourteen have 300 to 999 employees, accounting for 16%, and fifty-nine have more than 1,000 employees, accounting for 69%. There is also a variety of the companies' years of existence. Two companies are less than 3 years old; five companies are 3 to 5 years old; three companies are 6 to 10 years old, and the rest seventy-five companies are more than 10 years old. The business scope of these eighty-six manufacturing companies varies, namely general equipment manufacturing, special equipment manufacturing, transportation equipment manufacturing, electronics manufacturing, food manufacturing, etc., as shown in Figure 3.

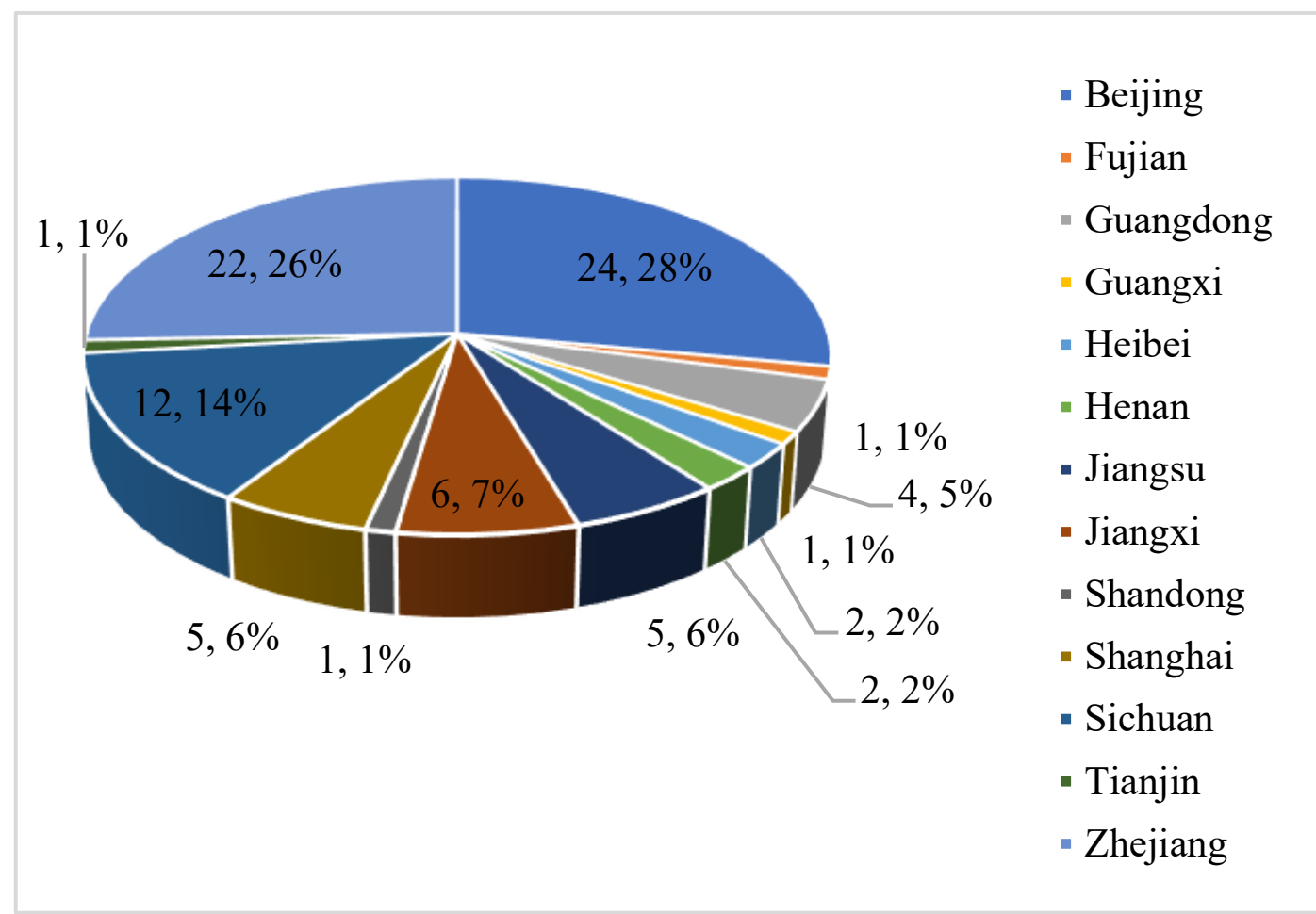

Fig 2. Regions of the companies

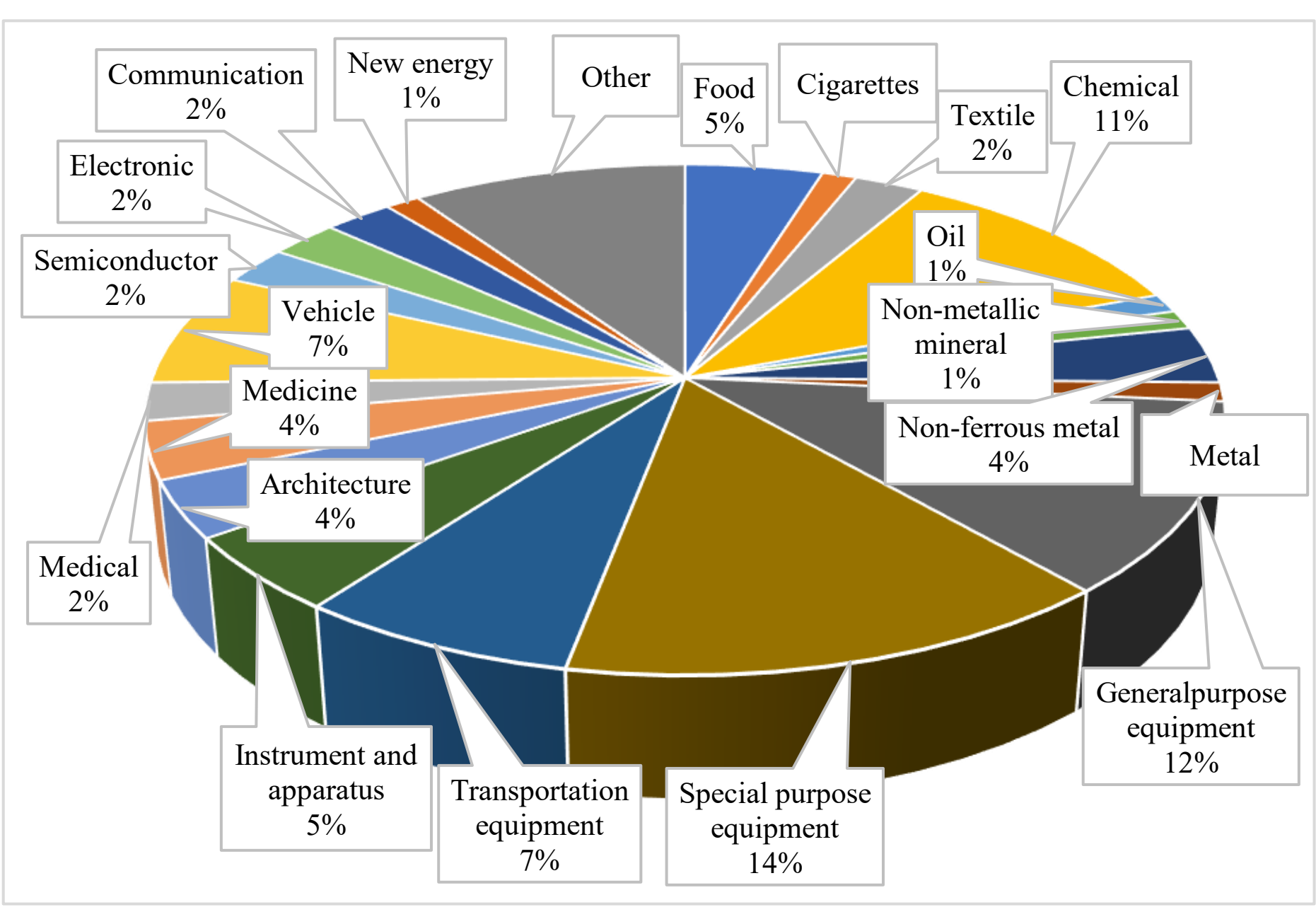

Fig 3. Business scope of the companies

As mentioned above, the positions and professions of the interviewees, as well as the regions, sizes, existence time and business scope of the companies they work for are diverse, which ensures the representativeness of the survey samples.

## 3.4 Reliability and Validity

We chose Partial Least Squares Structure Equation Modeling (PLS-SEM) to test our hypotheses, because PLS-SEM shows good efficiency dealing with small sample sizes (Reinartz et al., 2009), and our sample size satisfies the guidelines suggested by Marcoulides & Saunders (2006). The results of reliability and validity are demonstrated below.

### 3.4.1 Confirmatory Factor Analysis (CFA)

The PLS path model (drawn in Smart PLS software) after PLS Algorithm calculation with independent variables, dependent variables, relationship among variables and all indicators of variables are shown in Figure 4. Arrow directions indicate the reflective nature of variables.

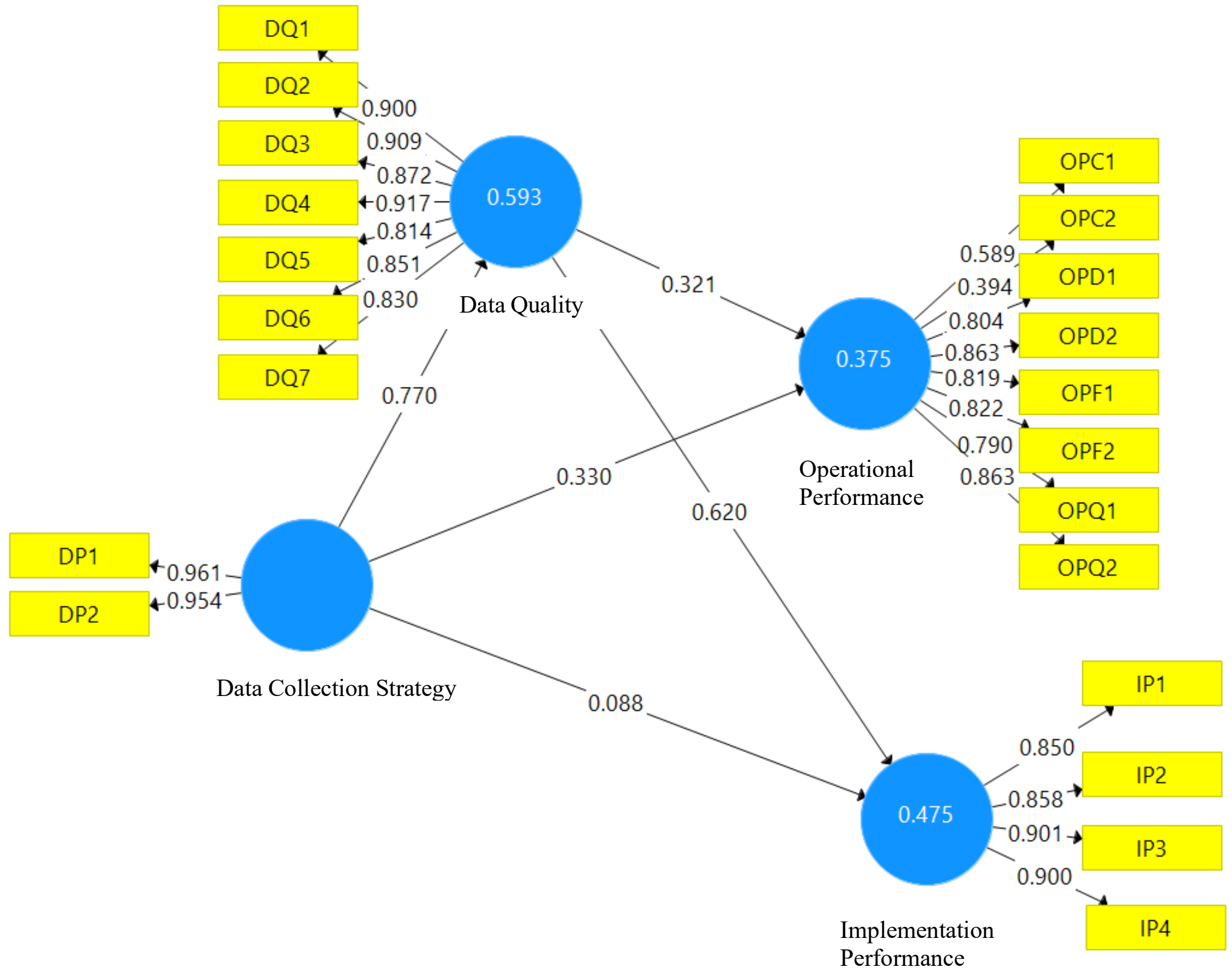

Fig. 4. PLS path model

In our analysis, operational performance and implementation performance are dependent variables and data collection strategy is an independent variable. The result shows that data collection strategy affects data quality, operational performance, and implementation performance, and the path coefficient are 0.770, 0.330, and 0.088, respectively. Data quality is a mediating variable, and it affects operational performance and implementation with path coefficient of 0.321 and 0.620, respectively.

### 3.4.2 Construct Reliability and Validity Metrics

The values of Cronbach's alpha, rho_A, composite reliability (CR), and average variance extracted (AVE) were computed to check the reliability and validity of model, as showing in Table2.

If the values of Cronbach's alpha are greater than 0.7 then the items are considered acceptable. An item with a Cronbach's alpha value of 0.4 to 0.7 should be removed if that increases the CR and AVE value more than threshold value. The value of Cronbach's alpha less than 0.4 shows that the item should be extracted (Chin 2010; Hair, et al. (2017). The values of Cronbach's alpha for data collection strategy, data quality, implementation performance, and operational performance are 0.909, 0.947, 0.901 and 0.888, respectively. All the values are greater than 0.7.

The value of CR is also calculated to check the reliability of constructs. The results show that CR values for data collection strategy, data quality, implementation performance, and operational performance are 0.956, 0.957, 0.930 and 0.912, respectively. All the CR values are greater than 0.7, which indicates that the model possesses acceptable level of reliability (Chin, 2010; Hair et al., 2021.) The AVE values of the latent variables were also computed and reflected in Table 3. The AVE value for data collection strategy, data quality, implementation performance, and operational performance are 0.917, 0.759, 0.770 and 0.576, respectively. All the values are greater than 0.5 which shows the convergent validity level is acceptable (Chin 2010; Hair et al., 2017).

Table 3. Construct Reliability and Validity Metrics

| | Cronbach's Alpha | rho_A | CR | AVE |
|---|---|---|---|---|
| **Data collection strategy** | 0.909 | 0.914 | 0.956 | 0.917 |
| **Data quality** | 0.947 | 0.948 | 0.957 | 0.759 |
| **Implementation performance** | 0.901 | 0.908 | 0.930 | 0.770 |
| **Operational performance** | 0.888 | 0.920 | 0.912 | 0.576 |

### 3.4.3 Discriminant Validity

Discriminant validity is demonstrated by evidence that measures of constructs that theoretically should not be highly related to each other are, in fact, not found to be highly correlated to each other (Hubley, 2014). Fornell-Larcker criterion is a widely used way to verify and confirm discriminant validity, and the diagonal values should be greater than non-diagonal values (Chin 2010; Hair et al. 2017). The results of Fornell-Larcker criterion are shown in Table 4, which indicate that no issue is found regarding discriminant validity.

Table 4. Fornell-Larcker Criterion

| | Data collection strategy | Data quality | Implementation performance | Operational performance |
|---|---|---|---|---|
| **Data collection strategy** | 0.957 | | | |
| **Data quality** | 0.770 | 0.871 | | |
| **Implementation performance** | 0.565 | 0.687 | 0.877 | |
| **Operational performance** | 0.577 | 0.575 | 0.529 | 0.759 |

Another criterion to check discriminant validity is the Heterotrait-Monotrait (HTMT) ratio. If the HTMT value is below 0.90, discriminant validity has been established between two reflective constructs (Hair et al., 2017). The results of HTMT of our model are shown in Table 5, and all the values are less than 0.9.

Table 5. HTMT ratio

| | Data collection strategy | Data quality | Implementation performance | Operational performance |
|---|---|---|---|---|
| **Data collection strategy** | | | | |
| **Data quality** | 0.829 | | | |
| **Implementation performance** | 0.618 | 0.740 | | |
| **Operational performance** | 0.619 | 0.610 | 0.577 | |

### 3.4.4 Outer Loadings

The values of factor loadings are considered to show the reliability of individual indicators of constructs. The value for factor loading should be more than 0.7 for acceptance. The values of the outer loading show the factor loading of all the constructs of latent variables. The results are shown in Table 6, in which the values for OPC1 and OPC2 are 0.589 and 0.394, respectively. These values can be dropped for getting improvement in final results.

Table 6. Outer loading

| Latent variable | Constructs | Factor loading |
|---|---|---|
| **Data collection strategy** | DP1 | 0.961 |
| | DP2 | 0.954 |
| **Data quality** | DQ1 | 0.900 |
| | DQ2 | 0.909 |
| | DQ3 | 0.872 |
| | DQ4 | 0.917 |
| | DQ5 | 0.814 |
| | DQ6 | 0.851 |
| | DQ7 | 0.830 |
| **Implementation performance** | IP1 | 0.850 |
| | IP2 | 0.858 |
| | IP3 | 0.901 |
| | IP4 | 0.900 |
| **Operational performance** | OPC1 | 0.589 |
| | OPC2 | 0.394 |
| | OPD1 | 0.804 |
| | OPD2 | 0.863 |
| | OPF1 | 0.819 |
| | OPF2 | 0.822 |
| | OPQ1 | 0.790 |
| | OPQ2 | 0.863 |

### 3.4.5 Evaluation of the Structural Model in PLS-SEM: Collinearity Assessment

Before drawing any conclusion, the structural model has to be properly evaluated. Variance inflation factor (VIF) values are used to check the potential issue of collinearity in the structural model. The outer VIF shows the severity of collinearity among items within a construct; additionally, the inner VIF shows the severity of collinearity among latent variables in the model. If the values of VIF are below 5, then it is considered that the issue of collinearity is not present among the variables (Hair et al., 2021).

The results of outer VIF value and inner VIF values are shown in Table 7 and 8, respectively. The results show that both the outer and inner VIF values are less than 5.

Table 7. Outer VIF values

| | VIF |
|---|---|
| **DP1** | 3.272 |
| **DP2** | 3.272 |
| **DQ1** | 4.014 |
| **DQ2** | 4.439 |
| **DQ3** | 3.935 |
| **DQ4** | 4.924 |
| **DQ5** | 2.904 |
| **DQ6** | 3.372 |
| **DQ7** | 2.879 |
| **IP1** | 2.400 |
| **IP2** | 2.486 |
| **IP3** | 3.368 |
| **IP4** | 3.242 |
| **OPC1** | 2.131 |
| **OPC2** | 1.852 |
| **OPD1** | 3.706 |
| **OPD2** | 4.203 |
| **OPF1** | 4.639 |
| **OPF2** | 4.087 |
| **OPQ1** | 2.704 |
| **OPQ2** | 3.324 |

Table 8. Inner VIF values

| | Data collection strategy | Data quality | Implementation performance | Operational performance |
|---|---|---|---|---|
| **Data collection strategy** | | 1.000 | 2.454 | 2.454 |
| **Data quality** | | | 2.454 | 2.454 |
| **Implementation performance** | | | | |
| **Operational performance** | | | | |

### 3.4.6 R Square

R-squared is a statistical measure of how close the data are to the fitted regression line. It is also known as the coefficient of determination, or the coefficient of multiple determination for multiple regression. The values of R square and R square adjusted for the latent variables are presented in Table 9. The values of R square for data quality, implementation performance, and operational performance are 0.593, 0.475 and 0.375, respectively, and the values of R square adjusted are 0.588, 0.463, and 0.360, respectively.

Table 9. R Square

| | R Square | R Square Adjusted |
|---|---|---|
| **Data quality** | 0.593 | 0.588 |
| **Implementation performance** | 0.475 | 0.463 |
| **Operational performance** | 0.375 | 0.360 |

### 3.4.7 F Square

The value of F square reflects the significance of one construct on another construct along with the degree of its effectiveness (Bhutta et al., 2019). F-Square is the change in R-Square when an exogenous variable is removed from the model. If the F Square value is equal to or below 0.02, then the effect size is considered small; if it's equal to or greater than 0.15, then the effect size is medium; if it's equal to or greater than 0.35, then the effect size is large (Cohen, 1988). The results of F square are shown in Table 10.

Table 10. F Square Values

| | Data collection strategy | Data quality | Implementation performance | Operational performance |
|---|---|---|---|---|
| **Data collection strategy** | | **1.454** | **0.006** | **0.071** |
| **Data quality** | | | **0.298** | **0.067** |
| **Implementation performance** | | | | |
| **Operational performance** | | | | |

### 3.4.8 Path Coefficients from Bootstrapping

Bootstrapping allows estimation of the sampling distribution of almost any statistic using random sampling methods (Varian, 2005), and it is any test or metric that uses random sampling with replacement, and falls under the broader class of resampling methods. Bootstrapping assigns measures of accuracy to sample estimates (Efron & Tibshirani, 1994). We used bootstrapping technique to check and test the significance of our model, by which the value of t-statistics reflects significance of path coefficients (Ringle, et al., 2015). According to Wong (2013), the critical t-value is 1.65 for a significance level of 10%, and 2.58 for a significance level of 1% (all two-tailed). The Table 11 shows the results of path coefficients. It can be seen that only the "Data collection strategy -> Implementation performance" linkage is not significant in our model.

Table 11. Path Coefficients from Bootstrapping

| | T Statistics (\|O/STDEV\|) | P Values |
|---|---|---|
| **Data process -> Data quality** | 14.153 | 0.000 |
| **Data process -> Implementation performance** | 0.633 | 0.527 |
| **Data process -> Operational performance** | 2.584 | 0.010 |
| **Data quality -> Implementation performance** | 4.720 | 0.000 |

## 4. Analyses and Results

### 4.1 Total Effects and Testing of Hypothesis

The results of the PLS-SEM analysis are presented in Table 12. The bootstrapping results show that all original sample values and sample means are positive. However, the path from data collection strategy → implementation performance ($t = 0.633$, $p = 0.572$) is not statistically significant, indicating that data collection strategy alone does not have a direct effect on implementation performance. In contrast, all other hypothesized paths are positive and significant. Specifically, data collection strategy has significant positive effects on data quality and operational performance, while data quality positively and significantly influences both implementation performance and operational performance. These findings suggest that while a well-designed data collection strategy contributes directly to better operational outcomes, its impact on implementation success may be realized indirectly through improved data quality.

Table 12. Total effects

| | Original Sample (O) | Sample Mean (M) | Standard Deviation (STDEV) | T Statistics (\|O/STDEV\|) | P Values | Decision |
|---|---|---|---|---|---|---|
| **Data process -> Data quality** | 0.770 | 0.768 | 0.054 | 14.153 | 0.000 | Supported |
| **Data process -> Implementation performance** | 0.088 | 0.091 | 0.138 | 0.633 | 0.527 | Not supported |
| **Data process -> Operational performance** | 0.330 | 0.338 | 0.128 | 2.584 | 0.010 | Supported |
| **Data quality -> Implementation performance** | 0.620 | 0.614 | 0.131 | 4.720 | 0.000 | Supported |
| **Data quality -> Operational performance** | 0.321 | 0.323 | 0.130 | 2.459 | 0.014 | Supported |

### 4.2 Mediation Analysis

To further examine the mediating role of data quality, we conducted mediation analysis using both total indirect effects and specific indirect effects. The results, shown in Tables 13 and 14, confirm that data quality acts as a significant mediator between data collection strategy and both performance outcomes. When the mediator is included, the indirect effects from data collection strategy to implementation performance and from data collection strategy to operational performance become significant. This indicates that data quality is a key mechanism through which data collection strategy enhances organizational performance.

Table 13. Total indirect effects

| | Original Sample (O) | Sample Mean (M) | Standard Deviation (STDEV) | T Statistics (\|O/STDEV\|) | P Values |
|---|---|---|---|---|---|
| **Data process -> Data quality** | | | | | |
| **Data process -> Implementation performance** | 0.477 | 0.472 | 0.109 | 4.366 | 0.000 |
| **Data process -> Operational performance** | 0.247 | 0.248 | 0.104 | 2.368 | 0.018 |
| **Data quality -> Implementation performance** | | | | | |
| **Data quality -> Operational performance** | | | | | |

Table 14. Specific indirect effects

| | Original Sample (O) | Sample Mean (M) | Standard Deviation (STDEV) | T Statistics (\|O/STDEV\|) | P Values |
|---|---|---|---|---|---|
| **Data process -> Data quality -> Implementation performance** | 0.477 | 0.472 | 0.109 | 4.366 | 0.000 |
| **Data process -> Data quality -> Operational performance** | 0.247 | 0.248 | 0.104 | 2.368 | 0.018 |

### 4.3 Summary of Hypothesis Testing:

Based on the model estimation and mediation analysis, all hypotheses are supported. The results confirm that: 1) Data collection strategy significantly influences implementation performance. 2) Data collection strategy significantly influences operational performance. 3) Data collection strategy significantly influences data quality. 4) Data quality significantly influences implementation performance. 5) Data quality significantly influences operational performance.

However, further examination of the mediation results reveals that the relationship between data collection strategy and implementation performance is fully mediated by data quality, as the direct effect is not significant once data quality is introduced. This means that an effective data collection strategy enhances implementation performance only through its improvement of data quality. In contrast, the relationship between data collection strategy and operational performance is partially mediated by data quality, as both direct and indirect effects are significant. Therefore, while high-quality data plays a central role in achieving superior performance, a well-structured data collection strategy also directly contributes to operational efficiency and flexibility.

In summary, these results highlight that data quality serves as a critical bridge between data management practices and digital transformation outcomes, demonstrating its dual role as both a direct performance driver and a mediator that channels the benefits of a robust data collection strategy.

## 5. Discussion and Implication

### 5.1 The Role of Data Collection Strategy

The findings confirm that a well-defined and standardized data collection strategy contributes significantly to both data quality and performance outcomes. This suggests that digital initiatives built upon structured and well-governed data collection processes are more likely to succeed. Standardization of data types, sources, and collection procedures ensures that the data feeding digital tools is consistent, complete, and relevant. Such structured approaches also facilitate system integration and reduce redundancies across business functions, creating a more reliable foundation for analytics, automation, and real-time decision-making.

However, the non-significant direct relationship between data collection strategy and implementation performance highlights an important nuance. While standardization and control in data collection create favorable conditions for digital projects, their benefits are not realized unless the collected data is of high quality. This implies that data collection efforts alone do not automatically translate into implementation success, therefore, they must be complemented by continuous validation, cleaning, and quality assurance mechanisms that ensure the usability of the data in digital contexts.

**5.2 The Central Role of Data Quality**

The study's results emphasize that data quality serves as a key mediating mechanism linking data collection strategy to both implementation and operational performance. High-quality data enhances the efficiency of implementation processes by reducing errors, rework, and integration challenges. It ensures that digital systems are built upon reliable and timely information, thereby improving adherence to project timelines and budgets. Moreover, high-quality data directly strengthens operational performance by supporting accurate forecasting, process optimization, and data-driven decision-making.

In line with prior research, this study confirms that data quality is not a static characteristic but the outcome of systematic governance and management practices. The complete mediation effect in implementation performance underscores that data quality is the channel through which data practices translate into tangible project success, while the partial mediation in operational performance highlights that both structured data processes and data quality jointly contribute to sustainable operational benefits.

**5.3 Theoretical Implications**

The findings extend existing theories of digital transformation and data governance in several ways. First, they provide empirical validation of the data foundation perspective, showing that data-related practices constitute a distinct dimension of readiness for digital transformation, alongside technological and organizational factors. Second, by distinguishing between data collection strategy and data quality, the study clarifies the causal mechanisms underlying digital success—moving beyond the assumption that data is merely an available input. Third, the identification of a full versus partial mediation pattern contributes to the understanding of how different performance outcomes are shaped at various stages of DTA, from implementation to post-adoption.

Overall, this research contributes to bridging the gap between data governance literature and digital transformation research, emphasizing that successful digitalization requires coordinated attention to both the generation and the management of data.

### 5.4 Managerial Implications

From a practical standpoint, the findings suggest several actionable insights for managers and practitioners. First, firms should design and formalize their data collection strategies early in the digital transformation process, ensuring clear definitions of data sources, formats, and responsibilities across departments. Second, organizations must invest in data quality management systems, including validation routines, metadata management, and cross-functional data governance structures, to guarantee that collected data remain reliable and usable throughout the project lifecycle. Third, managers should recognize that data quality is a performance enabler, not merely a technical issue. Allocating resources to continuous data improvement through auditing, standardization, and staff training can yield significant returns in both implementation efficiency and long-term operational performance.

## 6. Conclusion, Limitation, and Future Direction

### 6.1 Conclusion

This study sets out to explore how data collection strategy and data quality jointly influence the outcomes of DTA in the manufacturing context. Drawing on survey data and PLS-SEM analysis, the results provide strong empirical support for the proposed relationships. Specifically, data collection strategy significantly influences both data quality and operational performance, while its effect on implementation performance is fully mediated by data quality. Data quality, in turn, has significant positive effects on both implementation and operational performance, confirming its pivotal role as the bridge between upstream data practices and downstream digital outcomes.

These results reinforce the central argument proposed in the introduction: digital transformation success depends not only on technology and managerial readiness, but also on the quality and management of the underlying data. Although organizations increasingly recognize data as a strategic resource, this study empirically demonstrates that the way data is collected and managed determines the extent to which digital technologies can deliver their intended value.

### 6.2 Limitation

Despite its contributions, this study has several limitations that should be acknowledged.

First, the sample is limited to 86 manufacturing firms, which, although appropriate for PLS-SEM, may constrain the generalizability of results to other sectors such as logistics, services, or healthcare. Future studies could expand the empirical scope across different industries and institutional contexts. Second, the study relies on self-reported survey data, which may introduce subjectivity and common method bias. Although statistical tests indicated no major issues, future research could combine perceptual measures with objective performance data or archival datasets. Third, while the study focuses on two key data-related constructs, data collection strategy and data quality, but other dimensions of data management, such as data integration, governance structure, and metadata management, were not included. Incorporating these elements would allow a more comprehensive understanding of the data foundation underlying digital transformation.

### 6.3 Future Direction

This study demonstrates that data practices are not peripheral but foundational to digital success. By revealing how data collection strategy and data quality jointly shape implementation and operational performance, it underscores that effective data-driven decision-making depends not merely on gathering more data but on handling it with systematic governance mindset. Data quality serves as the critical link between data strategy and performance, suggesting that organizations must establish mechanisms to maintain and control data quality beyond the collection phase. Building on these insights, future research should expand from data collection toward a comprehensive data governance framework that integrates standardization, accountability, and lifecycle management. Such an approach can explain how firms institutionalize data quality assurance across the data lifecycle, spanning generation, storage, sharing, and utilization, to enable scalable, intelligent, and reliable operations in the era of Industry 4.0.

**Reference:**

Abbasi, A., Sarker, S., & Chiang, R. H. (2016). Big data research in information systems: Toward an inclusive research agenda. *Journal of the association for information systems*, 17(2), 3.

Abraham, R., Schneider, J., & Vom Brocke, J. (2019). Data governance: A conceptual framework, structured review, and research agenda. *International Journal of Information Management*, 49, 424–438.

Aldossari, S., Mokhtar, U. A., & Abdul Ghani, A. T. (2023). Factor influencing the adoption of Big Data Analytics: A systematic literature and experts review. *Sage Open*, *13*(4), 21582440231217902.

Akyuz, G. A., & Rehan, M. (2009). Requirements for forming an 'e-supply chain'. *International Journal of Production Research*, 47(12), 3265-3287.

Ballou, D. P., & Pazer, H. L. (1985). Modeling data and process quality in multi-input, multi-output information systems. *Management Science*, 31(2), 150–162.

Bandara, W., Gable, G. G., & Rosemann, M. (2005). Factors and measures of business process modelling: model building through a multiple case study. *European Journal of Information Systems*, 14(4), 347–360.

Barrett, M., Davidson, E., Prabhu, J., & Vargo, S. L. (2015). Service innovation in the digital age. *MIS Quarterly*, 39(1), 135–154.

Batini, C., Cappiello, C., Francalanci, C., & Maurino, A. (2009). Methodologies for data quality assessment and improvement. *ACM computing surveys (CSUR)*, *41*(3), 1-52.

Bhutta, E., Kausar, S., & Rehman, A. (2019). Factors affecting the performance of market committees in Punjab, Pakistan: An empirical assessment of performance through SMART PLS mediation analysis. *Journal of Agricultural Research*, 57(3).

Bharadwaj, A., El Sawy, O. A., Pavlou, P. A., & Venkatraman, N. V. (2013). Digital business strategy: Toward a next generation of insights. *MIS Quarterly*, 37(2), 471–482.

Bilsborrow, R. E. (2016). Concepts, definitions and data collection approaches. In *International handbook of migration and population distribution* (pp. 109-156). Dordrecht: Springer Netherlands.

Blichfeldt, H., & Faullant, R. (2021). Performance effects of DTA and product & service innovation–A process-industry perspective. *Technovation*, *105*, 102275.

Bosu, M. F., & MacDonell, S. G. (2013, April). Data quality in empirical software engineering: a targeted review. In *Proceedings of the 17th International Conference on Evaluation and Assessment in Software Engineering* (pp. 171-176).

Cagliano, R., Canterino, F., Longoni, A., & Bartezzaghi, E. (2019). The interplay between smart manufacturing technologies and work organization: the role of technological complexity. *International*

*Journal of Operations & Production Management*.

Cheng, Y., Farooq, S., & Jajja, M. S. S. (2020). Does plant role moderate relationship between internal manufacturing network integration, external supply chain integration, operational performance in manufacturing network? *Journal of Manufacturing Technology Management*.

Chien, S. W., Lin, H. C., & Shih, C. T. (2014). A moderated mediation study: Cohesion linking centrifugal and centripetal forces to ERP implementation performance. *International Journal of Production Economics*, 158, 1–8.

Cifone, F. D., Hoberg, K., Holweg, M., & Staudacher, A. P. (2021). 'Lean 4.0': How can digital technologies support lean practices? *International Journal of Production Economics*, 241, 108258.

Cohen, J. (1988). *Statistical power analysis for the behavioral sciences* (2nd ed.). Hillsdale, NJ: Erlbaum.

Efron, B., & Tibshirani, R. J. (1994). *An introduction to the bootstrap*. Boca Raton, FL: CRC Press.

El Sawy, O. A., Kræmmergaard, P., Amsinck, H., & Vinther, A. L. (2016). How LEGO built the foundations and enterprise capabilities for digital leadership. *MIS Quarterly Executive*, 15(2).

Esther Shein. (2024, December 16). 8 reasons why digital transformations still fail. *CIO*. https://www.cio.com/article/228268/12-reasons-why-digital-transformations-fail.html

Evans, P. (2006). Scaling and assessment of data quality. *Biological crystallography*, *62*(1), 72-82.

Gillani, F., Chatha, K. A., Jajja, M. S. S., & Farooq, S. (2020). Implementation of digital manufacturing technologies: Antecedents and consequences. *International Journal of Production Economics*, 229, 107748.

Gökalp, M. O., Gökalp, E., Kayabay, K., Koçyiğit, A., & Eren, P. E. (2021). Data-driven manufacturing: An assessment model for data science maturity. *Journal of Manufacturing Systems*, 60, 527–546.

Hair, J. F., Hult, G. T. M., Ringle, C. M., and Sarstedt, M. (2017). *A Primer on Partial Least Squares Structural Equation Modeling (PLS-SEM)*., 2nd Ed., Thousand Oakes, CA: Sage.

Hair Jr, J. F., Hult, G. T. M., Ringle, C. M., & Sarstedt, M. (2021). *A Primer on Partial Least Squares Structural Equation Modeling (PLS-SEM)*. Sage Publications.

Hazen, B. T., Boone, C. A., Ezell, J. D., & Jones-Farmer, L. A. (2014). Data quality for data science, predictive analytics, and big data in supply chain management: An introduction to the problem and suggestions for research and applications. *International Journal of Production Economics*, 154, 72-80.

Heinen, J. J., & Hoberg, K. (2019). Assessing the potential of additive manufacturing for the provision of spare parts. *Journal of Operations Management*, 65(8), 810–826.

Holmström, J., Holweg, M., Lawson, B., Pil, F. K., & Wagner, S. M. (2019). The digitalization of operations and supply chain management: Theoretical and methodological implications. *Journal of Operations Management*, 65(8), 728–734.

Holotiuk, F., & Beimborn, D. (2017). Critical success factors of digital business strategy. In *Wirtschaftsinformatik Conference*, St. Gallen, Switzerland. AIS Electronic Library, 991–1005.

Hubley, A. M. (2014). Discriminant validity. In A. C. Michalos (Ed.), *Encyclopedia of Quality of Life and Well-Being Research*. Springer. https://doi.org/10.1007/978-94-007-0753-5_751

Kiraz, A., Canpolat, O., Özkurt, C., & Taşkın, H. (2020). Analysis of the factors affecting the Industry 4.0 tendency with the structural equation model and an application. *Computers & Industrial Engineering*, *150*, 106911.

Lee, J., Bagheri, B., & Kao, H. A. (2015). A cyber-physical systems architecture for industry 4.0-based manufacturing systems. *Manufacturing Letters*, 3, 18–23.

Li, L., Su, F., Zhang, W., & Mao, J. Y. (2018). Digital transformation by SME entrepreneurs: A capability perspective. *Information Systems Journal*, *28*(6), 1129-1157.

Marcon, É., Soliman, M., Gerstlberger, W., & Frank, A. G. (2022). Sociotechnical factors and Industry 4.0: an integrative perspective for the adoption of smart manufacturing technologies. *Journal of Manufacturing Technology Management*, *33*(2), 259-286.

Marcoulides, G. A., & Saunders, C. (2006). Editor's comments: PLS: a silver bullet? *MIS Quarterly*, iii–ix.

Morakanyane, R., O'Reilly, P., McAvoy, J., & Grace, A. (2020). Determining digital transformation success factors. In *Proceedings of the 53rd Hawaii International Conference on System Sciences*.

Möller, D. P. (2020). Introduction to Digital Transformation. In *Cybersecurity in Digital Transformation: Scope and Applications* (pp. 1-10). Cham: Springer International Publishing.

Ofner, M. H., Otto, B., & Österle, H. (2012). Integrating a data quality perspective into business process management. *Business Process Management Journal*.

Ram, J., Corkindale, D., & Wu, M. L. (2013). Implementation critical success factors (CSFs) for ERP: Do they contribute to implementation success and post-implementation performance? *International Journal of Production Economics*, 144(1), 157–174.

Ramesh, N., & Delen, D. (2021). Digital transformation: How to beat the 90% failure rate?. *IEEE engineering management review*, *49*(3), 22-25.

Rasouli, M. R., Trienekens, J. J., Kusters, R. J., & Grefen, P. W. (2016). Information governance requirements in dynamic business networking. *Industrial Management & Data Systems*.

Reinartz, W., Haenlein, M., & Henseler, J. (2009). An empirical comparison of the efficacy of covariance-based and variance-based SEM. *International Journal of Research in Marketing*, 26(4),

332–344.

Ringle, C., Da Silva, D., & Bido, D. (2015). Structural equation modeling with the SmartPLS. *Brazilian Journal of Marketing*, 13(2).

Roscoe, S., Cousins, P. D., & Handfield, R. (2019). The microfoundations of an operational capability in digital manufacturing. *Journal of Operations Management*, 65(8), 774–793.

Sivarajah, U., Kamal, M. M., Irani, Z., & Weerakkody, V. (2017). Critical analysis of Big Data challenges and analytical methods. *Journal of business research*, *70*, 263-286.

Skare, M., & Soriano, D. R. (2021). How globalization is changing DTA: An international perspective. *Journal of Innovation & Knowledge*, *6*(4), 222-233.

Svahn, F., Mathiassen, L., & Lindgren, R. (2017). Embracing digital innovation in incumbent firms: How Volvo Cars managed competing concerns. *MIS Quarterly*, 41(1), 239–253.

Tan, C. L., & Vonderembse, M. A. (2006). Mediating effects of computer-aided design usage: From concurrent engineering to product development performance. *Journal of Operations Management*, 24(5), 494–510.

Tilson, D., Lyytinen, K., & Sørensen, C. (2010). Research commentary—Digital infrastructures: The missing IS research agenda. *Information Systems Research*, 21(4), 748–759.

Tortorella, G. L., Fogliatto, F. S., Esposto, K. F., Mac Cawley Vergara, A., Vassolo, R., Tlapa Mendoza, D., & Narayanamurthy, G. (2022). Measuring the effect of Healthcare 4.0 implementation on hospitals' performance. *Production Planning & Control*, 33(4), 386-401.

Varian, H. (2005). Bootstrap tutorial. *Mathematica Journal*, 9(4), 768–775.

Visich, J. K., Li, S., Khumawala, B. M., & Reyes, P. M. (2009). Empirical evidence of RFID impacts on supply chain performance. *International Journal of Operations & Production Management*.

Fan, W., & Geerts, F. (2012). *Foundations of data quality management*. Morgan & Claypool Publishers.

Wang, R. Y., & Strong, D. M. (1996). Beyond accuracy: What data quality means to data consumers. *Journal of management information systems*, 12(4), 5-33.

Wang, Y., Kung, L., & Byrd, T. A. (2018). Big data analytics: Understanding its capabilities and potential benefits for healthcare organizations. *Technological forecasting and social change*, *126*, 3-13.

Wixom, B. H., & Watson, H. J. (2001). An empirical investigation of the factors affecting data warehousing success. *MIS Quarterly*, 25(1), 17–41.

Wong, K. K. K. (2013). Partial least squares structural equation modeling (PLS-SEM) techniques using SmartPLS. *Marketing Bulletin*, 24(1), 1–32.

Xu, H., Nord, J. H., Brown, N., & Nord, G. D. (2002). Data quality issues in implementing an ERP. *Industrial Management & Data Systems*.

Yang, M., Fu, M., & Zhang, Z. (2021). The adoption of digital technologies in supply chains: Drivers, process and impact. *Technological Forecasting and Social Change*, *169*, 120795.

Yoo, Y., Boland Jr, R. J., Lyytinen, K., & Majchrzak, A. (2012). Organizing for innovation in the digitized world. *Organization Science*, 23(5), 1398–1408.

Zonta, T., Da Costa, C. A., da Rosa Righi, R., de Lima, M. J., Da Trindade, E. S., & Li, G. P. (2020). Predictive maintenance in the Industry 4.0: A systematic literature review. *Computers & Industrial Engineering*, 150, 106889.

**Appendix:**

Questionnaire: Data Practices and Digital Technology Adoption in Manufacturing Firms

Part I: Basic Information

1. Your job position: ________________________________________
2. Company location: __________________________ Province/City
3. Years since company establishment: __________________________
4. Total number of employees: __________________________
5. Main business/industry sector: __________________________
6. Which digital technologies has your company implemented, and in which business areas?

---

Part II: Data Collection Process

*(Please indicate the extent to which you agree with each statement.)*

Scale: 1 = Strongly disagree, 5 = Strongly agree

1. Before implementing digital technologies, our company's existing processes were standardized and clearly defined in terms of what data types and formats needed to be collected.
2. Before implementing digital technologies, our company had standardized data collection procedures, ensuring that data were collected consistently each time.

---

Part III: Data Quality

*("Data" here refers to the data generated after the implementation of digital technologies.)*

Please rate your level of agreement with the following statements:

Scale: 1 = Strongly disagree, 5 = Strongly agree

1. The data generated during the implementation of digital technologies are accurate, objective, and reliable.
2. The data generated during the implementation of digital technologies are updated in real time.
3. The data generated during the implementation of digital technologies are complete and comprehensive, providing all necessary information.
4. The data are expressed in appropriate formats and are clear, unambiguous, and easy to understand.
5. The data formats remain consistent throughout the implementation process.
6. The data can be easily and quickly retrieved and accessed.
7. The data are helpful and useful for the company's current business operations.

---

Part IV: Implementation Performance

*(Please indicate your level of agreement with each statement.)*

1. The implementation of digital technologies in our company was completed on schedule.
2. The implementation of digital technologies in our company was completed within budget.
3. The implementation outcomes of digital technologies met our expectations.
4. Our company is satisfied with the implementation of digital technologies.

---

Part V: Operational Performance

*(Compared to your main competitors, how would you rate your company's performance after implementing digital technologies?)*

Scale: 1 = Much worse, 5 = Much better

1. Ordering cost (the cost incurred each time an order is placed with a supplier)

2. Unit production cost
3. Quality consistency (the extent to which products or services meet planned standards)
4. Product quality and reliability (the degree to which products are reliable)
5. Capacity flexibility (the ability to quickly respond to market demand changes and adjust production volume in a short time)
6. Product mix flexibility (the ability to produce a variety of products at low switching cost)
7. Delivery speed (the ability to deliver products to customers on time)
8. Delivery reliability (the ability to ensure reliable and dependable deliveries to customers)